\documentclass[journal]{IEEEtran}
\usepackage{ifpdf}
\usepackage{cite}

\ifCLASSINFOpdf
  \usepackage[pdftex]{graphicx}
  \DeclareGraphicsExtensions{.pdf,.jpeg,.png}
\else
\fi
\begin{document}
%
\title{AM to PM conversion of linear filters}
%
%
%

\author{Magnus~Danielson,~\IEEEmembership{Member,~IEEE}}

%
%

\markboth{Journal of \LaTeX\ Class Files,~Vol.~14, No.~8, August~2015}%
{Shell \MakeLowercase{\textit{et al.}}: Bare Demo of IEEEtran.cls for IEEE Journals}
%



\maketitle

\begin{abstract}
  The conversion between amplitude modulation and phase modulation as a
  modulated signal goes thrugh a filter is analyzed. The difference in how
  the modulated sideband ampitude experience the filter, and how AM and PM has
  opposite signs for one of their sidebands interact. The conversion between
  AM and PM is modelled, providing a scatter model and evaluation of two
  functions based on the linear filters transfer function. The system bandwidth
  effects is analyzed and rule of thumb developed to ensure AM and PM isolation.
\end{abstract}


%
\IEEEpeerreviewmaketitle

\section{Introduction}
%
%
%
%
\IEEEPARstart{T}{he} amplitude modulation to phase modulation interaction of
system components can severly limit the achived phase noise performance of
high performance systems. This paper aims to facilitate a simplified approach
to the analysing of how such conversion can occur in linear filters.

The concern for both AM and PM noise is illustrated in \cite{NIST1139}, which
also touches on the subject of AM to PM noise conversion and therefore the need
to keep AM noise within constraints.
While it covers behavior of non-linear amplifier circuits, the focus of the
present article is on strictly linear behavior and a simplified approach to
analyze this in order to assist on system design.

\section{Modulation basics}
For amplitude modulation, the sidebands can be expressed using classic
trigonometry \cite{beta}
\begin{equation}
\begin{array}{rl}
X(t)& = A(1+a \cos \omega_m t)\cos \omega_ct\\
& = A\frac{a}{2} \cos (\omega_c-\omega_m)t + A \cos \omega_ct\\& + A\frac{a}{2} \cos (\omega_c+\omega_m)t
\end{array}
\end{equation}
Where $A$ is the amplitude, $a$ is the amplitude modulation index, $\omega_c$
the angular frequency of the carrier and $\omega_m$ the angular frequency of the
modulation.

Similarly, for phase modulation we can write from Bessel function \cite{beta}
assuming higher order terms insignificant:
\begin{equation}
\begin{array}{rl}
X(t)&= A\cos p\sin \omega_mt + \omega_ct\\
    &= AJ_{-1}(p)\cos (\omega_c-\omega_m)t + AJ_0(p) \cos \omega_ct\\&+ AJ_1(p)\cos (\omega_c+\omega_m)
\end{array}
\end{equation}
The following analysis assumes low modulation index, such that for phase
modulation effects on carrier strength can be considered insignificant. Thus,
can quadratic or higher orders of the Bessel polynoms be cancelled out and the
omission of second-degree or higher order sidebands also be dropped safely.
The remaining approximation thus becomes:
\begin{equation}
X(t) = -A\frac{p}{2}\cos (\omega_c-\omega_m)t + A \cos \omega_ct + A\frac{p}{2}\cos (\omega_c+\omega_m)
\end{equation}

\section{Common AM and PM model}

Giving the similarities of how AM and PM creates sidebands, only differing in
the sign of the lower sideband, we can create a model for simultanous AM and PM
where by the amplitude of the lower sides band LSB $A_{LSB}$ and upper side band
USB $A_{USB}$ can be expressed directly in the form of the respective modulatio
index $a$ and $p$ as well as the carrier amplitude $A$.
\begin{eqnarray}
A_{LSB} &=& A\frac{a}{2} - A\frac{p}{2}\\
A_{USB} &=& A\frac{a}{2} + A\frac{p}{2}
\end{eqnarray}
similarly can the respective modulation indexes be expressed in terms of the
LSB and USB amplitudes and carrier amplitude for the same modulation frequency:
\begin{eqnarray}
a &=& \frac{A_{USB} + A_{LSB}}{A}\\
p &=& \frac{A_{USB} - A_{LSB}}{A}
\end{eqnarray}
This thus represents an ortogonal linear transformation between either two
sidebands of a carrier or the AM and PM modulations of that carrier.
Depending on what we do we view it in either of these views and as long as we
have both amplitudes we can transform to the other view.

\section{AM and PM in linear filter}

Consider that we have a linear filter $H(s)$ and a signal that has amplitude and
phase modulation, how can the modulations be considered to behave? To answer
this question, one first need to convert the modulation indexes into the
relative strengths of the sidebands $A_{LSB}$ and $A_{USB}$.
With these, the filters response to the sideband frequencies produces the
output strengths $A_{LSB}'$ and $A_{USB}'$ while the carrier produces the output
strength $A_C'$ which can the be recalculated into the modulation indexes $a'$
and $p'$. This thus becomes
\begin{eqnarray}
\omega_u &=& \omega_c+\omega_m\\
\omega_l &=& \omega_c-\omega_m\\
A_{LSB} &=& A\frac{a-p}{2}\\
A_{USB} &=& A\frac{a+p}{2}\\
A_{LSB}' &=& H(j\omega_l)A_{LSB}\\
A_{C}' &=& H(j\omega_c)A\\
A_{USB'} &=& H(j\omega_u)A_{USB}\\
a' &=& \frac{A_{USB}'+A_{LSB}'}{A'}\\
p' &=& \frac{A_{USB}'-A_{LSB}'}{A'}
\end{eqnarray}

reducing into

\begin{eqnarray}
a' &=& \frac{H(j\omega_u)}{H(j\omega_c)}\frac{a+p}{2} + \frac{H(j\omega_l)}{H(j\omega_c)}\frac{a-p}{2}\\
p' &=& \frac{H(j\omega_u)}{H(j\omega_c)}\frac{a+p}{2} - \frac{H(j\omega_l)}{H(j\omega_c)}\frac{a-p}{2}\\
a' &=& \frac{H(j\omega_u) + H(j\omega_l)}{2H(j\omega_c)}a + \frac{H(j\omega_u) - H(j\omega_l)}{2H(j\omega_c)}p\\
p' &=& \frac{H(j\omega_u) - H(j\omega_l)}{2H(j\omega_c)}a + \frac{H(j\omega_u) + H(j\omega_l)}{2H(j\omega_c)}p
\end{eqnarray}

The last formulation clearly indicate that AM to AM and PM to PM conversion
depends on the common sideband response where as the AM to PM and PM to AM
conversion depends on the differential of the sideband response. It behaves as
a scattering matrix of a linear system. Thus can further simplification be
performed by defining the common and differential responses.
\begin{eqnarray}
H_c(\omega_l,\omega_u) &=& \frac{H(j\omega_u)+H(j\omega_l)}{2H(j\omega_c)}\\
H_d(\omega_l,\omega_u) &=& \frac{H(j\omega_u)-H(j\omega_l)}{2H(j\omega_c)}\\
a' &=& H_c(\omega_l,\omega_u)a + H_d(j\omega_l,j\omega_u)p\\
p' &=& H_d(\omega_l,\omega_u)a + H_c(j\omega_l,j\omega_u)p\\
\left[\begin{array}{c}a'\\p'\end{array}\right] &=& \left[\begin{array}{cc}H_c & H_d\\H_d&H_c\end{array}\right] \left[\begin{array}{c}a\\p\end{array}\right]
\end{eqnarray}

The AM to PM and PM to AM leakage depends on the difference in response, and
is equal. The AM to AM and PM to PM depends on the common response. However,
both also depends on the damping and is both sidebands sufficiently damped,
there will be significan damping of both. A perfectly balanced filter response
will cancel cross-modulation without high reduction of modulation transfer.

\section{1 pole lowpass filter}

To illustrate the effect, consider a 1 pole lowpass filter

\begin{equation}
H(s) = \frac{-\omega_0}{s-\omega_0}
\end{equation}

To analyze is, it gets inserted into the two formulas resulting in

\begin{eqnarray}
H_c(\omega_l,\omega_u) &=& \frac{\omega_c^2-\omega_0^2+j2\omega_0\omega_c}{\omega_0^2 - \omega_c^2 + \omega_m^2 -j2\omega_0\omega_c}\\
H_d(\omega_l,\omega_u) &=& \frac{\omega_c\omega_m + j\omega_0\omega_m}{\omega_0^2 - \omega_c^2 + \omega_m^2 -j2\omega_0\omega_c}
\end{eqnarray}

As these is evaluated for different relationships of carrier and frequency
relationships, assuming modulation frequency is low compared to carrier
frequency:

\begin{equation}
\begin{array}{rccc}
Condition & \left|H_c\right| & \left|H_d\right| & \left|A_C'\right| \\
\omega_0 >> \omega_c & 1 & 0 & A\\
\omega_0 = k\omega_c & 1 & \frac{\omega_m}{k\omega_0} = \frac{f_m}{kf_o} & A\\
\omega_0 = \omega_c & 1 & \frac{\omega_m}{\sqrt{2}\omega_c} = \frac{f_m}{\sqrt{2}f_c} & \frac{A}{\sqrt{2}}\\
\omega_0 << \omega_c & 1 & \frac{\omega_m}{\omega_c} = \frac{f_m}{f_c} & 0
\end{array}
\end{equation}

The $\left|H_c\right|$ response is essentially that of all pass for all
conditions, but notice how the amplitude of carrier reduces to reflex the
low-pass filter itself, so the AM to AM and PM to PM conversions both
experience the same pass action that we expect. Further notice how the
$\left|H_d\right|$ reflect a high-pass filter as scaled by $f_m/f_c$ factor.

The first condition reflect the situation where the carrier and sidebands is
well within the pass frequency of the filter, and during this condition there
is no cross-talk nor alteration of the carrier and thus the AM to AM and
PM to PM conversion works as expected.

The third condition reflect the situation where the carrier matches that of the
filter bandwidth, at which there is a 3~dB loss of amplitude, and there is a
cross-talk proportional to the $f_m/f_c$ ratio. Thus, a filter set at the
carrier frequency will provide AM to PM conversion that increases proporional
with the modulation frequency. Thus, far-out AM noise can convert to far-out
PM noise. A 100 kHz modulation of a 10 MHz source would have a -43~dB conversion
strength.

The second condition was added to reflect that increasing the system bandwidth
to be some ratio $k$ times the carrier would allow for additional isolation,
which can be readily seen, such that $k=10$ would provide about -60~dB
conversion strengt of AM to PM for the same 100 kHz of 10 MHz. This allows for
a simple rule of thumb approach to ensure AM to PM conversion does not impact
system performance.

The fourth condition was added to reflect the extreme case where system
bandwidth is far below that of the carrier. At this condition, there is AM to
PM conversion, but on the other hand the damping of carrier is significant such
that the gain goes towards zero and the carrier and sidebands is replaced by
thermal noise.

%

\section{Conclusion}
The fundamental approach to analyze AM to AM, PM to PM as well as cross-talk of
AM to PM and PM to AM has been done and can be summarized by the response of
the two $H_C(\omega_c, \omega_m)$ and $H_D(\omega_c, \omega_m)$ responses, that
can analyze the effect of any linear system $H(s)$.

For low-pass filter action, AM to PM cross-talk is found whenever the carrier
frequency is near or beyond the system bandwidth, but not when the carrier
frequency is much less than the system bandwidth. The cross-talk increases with
modulation frequency and is proportional to the ratio $f_m/f_c$. By letting the
system bandwidth be scaled by the factor of $k$ up from the carrier frequency,
as defined by $f_0 = k*f_c$, the coupling factor approximate $f_m/(k f_c)$ which
can be used as a rule of thumb to ensure enough isolation between AM and PM for
far-out noise.

\section*{Acknowledgment}

The authors would like to thank Craig Nelson, Archita Hati and David Howe, all
at NIST Boulder, as well as Bob Camp, Prof. Enrico Rubiola and Prof. Francois
Vernotte for their continuous support and inspiration.
The basic idea for this work comes out of Craig Nelson's Phase-noise tutorial
as being held at EFTF and IFCS conferences, where the different signs of
sideband amplitudes for AM and PM where illustrated.

\ifCLASSOPTIONcaptionsoff
  \newpage
\fi



%

%
\begin{IEEEbiography}[{\includegraphics[width=1in,height=1.25in,clip,keepaspectratio]{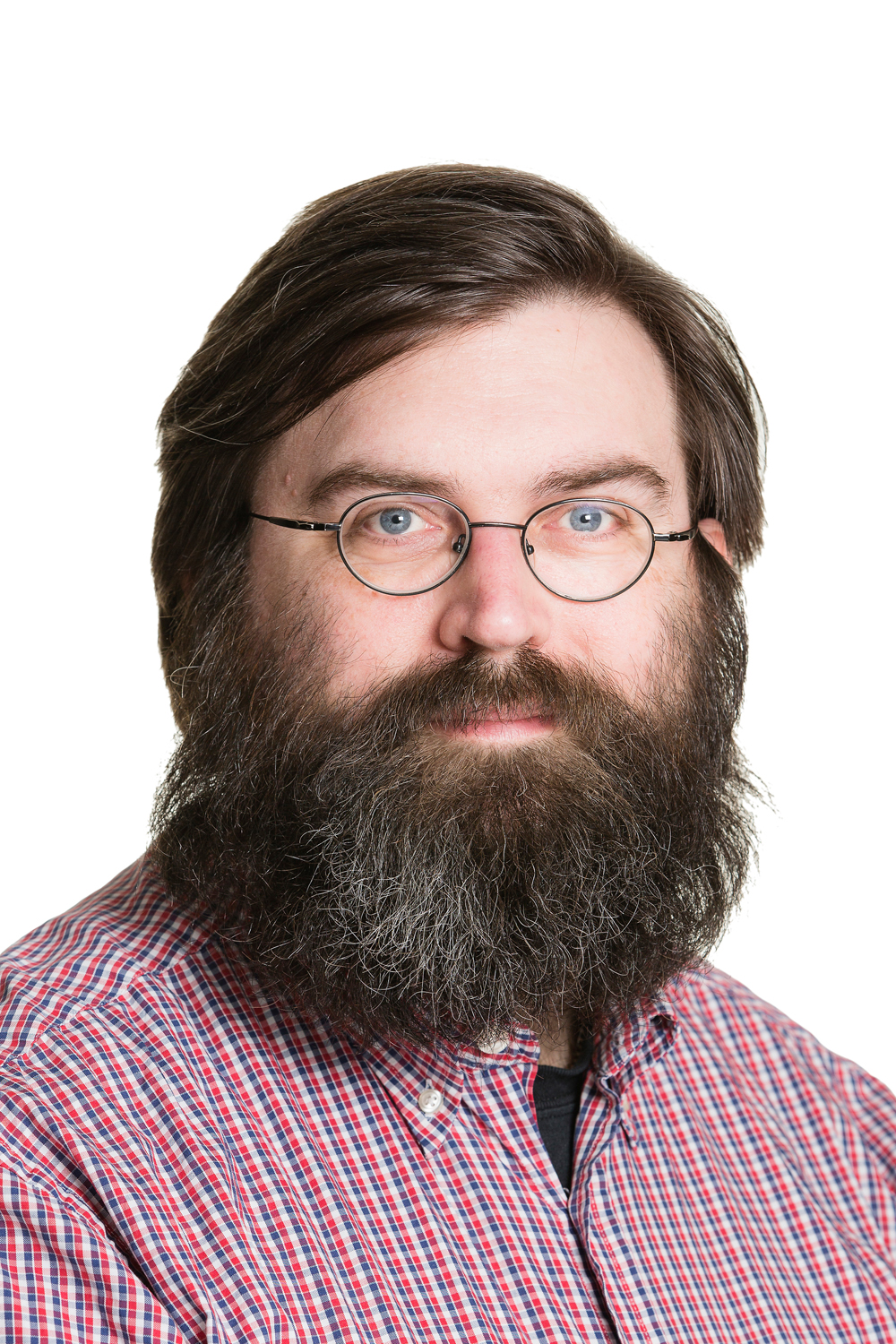}}]{Magnus Danielson}
  was born in Danderyd, Sweden, on April 27, 1971. Parallel to his studies he worked on
  professinal audio systems. His EE studies at Royal Institute of Technology
  started in 1993, but got interrupted as he got hired by the Department of
  Teleinformatics in 1994 where he worked as a Research Engineer for three
  years. In 1997 he followed a reasearch group to their start-up, and he has
  been with Net Insight since. At Net Insight he works as Senior System
  Architect, overseeing synchronization, timetransfer, audio transports,
  large scale media systems, protocol design, EMC and high speed signal
  integrity, standardization at ETSI. He has contributed to 18 patents, mainly
  focusing on synchronization, time-transfer and switching mechanisms.
  He is a hobby researcher in the time and frequency field, including authoring
  Allan variance article on Wikipedia. He also works with PNT issues,
  GPS/GNSS and related, and has presented before PNT Advisory Board on the
  use and operational issues of GPS in the commercial field. In his spare time
  he is a ham with call sign SA0MAD, contributing to rewriting the swedish ham
  educational material.
  He is a member of IEEE, AES, SMPTE and ION.
\end{IEEEbiography}




\end{document}